\documentclass{KapProc}

\normallatexbib

\usepackage{graphicx}

\begin{document}

\articletitle{Spin and charge order in the vortex lattice of the
cuprates: experiment and theory}

\author{Subir Sachdev}
\affil{Department of Physics, Yale University,\\
P.O. Box 208120, New Haven CT 06520-8120, USA} \email{Email:
subir.sachdev@yale.edu;
Web: http://pantheon.yale.edu/\~\/subir\\
Date: February 15, 2002; updated October 15, 2003}

\chaptitlerunninghead{Vortices in the cuprate superconductors}

\begin{abstract}
I summarize recent results, obtained with E.~Demler, K.~Park,
A.~Pol\-kov\-ni\-kov, M.~Vojta, and Y.~Zhang, on spin and charge
correlations near a magnetic quantum phase transition in the
cuprates. STM experiments on slightly overdoped
Bi$_2$Sr$_2$CaCu$_2$O$_{8+\delta}$ (J.~E.~Hoffman {\em et al.},
Science {\bf 295}, 466 (2002)) are consistent with the nucleation
of static charge order coexisting with dynamic spin correlations
around vortices, and neutron scattering experiments have measured
the magnetic field dependence of static spin order in the
underdoped regime in La$_{2-\delta}$Sr$_{\delta}$CuO$_4$ (B.~Lake
{\em et al.}, Nature {\bf 415}, 299 (2002)) and LaCuO$_{4+y}$
(B.~Khaykovich {\em et al.}, Phys. Rev. B {\bf 66}, 014528 (2002)
). Our predictions provide a semi-quantitative description of
these observations, with only a single parameter measuring
distance from the quantum critical point changing with doping
level. These results suggest that a common theory of competing
spin, charge and superconducting orders provides a unified
description of all the cuprates.
\end{abstract}

~\\~\\
 \noindent {\tt  Mexican Meeting on Mathematical and
Ex\-peri\-men\-tal Physics, Col\-legio Nacional, Mexico City,
September 2001. Published in {\em Dev\-elopments in Mathematical
and Experimental Physics, Volume B: Statistical Physics and
Beyond}, A. Macias, F. Uribe, and E. Diaz eds, Kluwer Academic,
New York (2002); }

\newpage

\section*{Introduction}
Three recent experiments \cite{seamus,lake2,boris} have shed new
light on the spin and charge density wave collective modes of the
cuprate superconductors. This article will summarize the main
results of our theory \cite{vs,prl,kwon,sns,pphmf,prb,rc} of these
collective modes in the vicinity of a quantum phase transition
between two superconducting states, only one of which has static,
long-range, spin density wave order: we will recall our main
predictions, and discuss further experimental tests. we will also
connect our theory to these experiments. In particular, we
suggested \cite{kwon} that static charge order should coexist with
dynamic spin-gap fluctuations around the vortex cores in the
cuprate superconductors, as may have been been observed in
\cite{seamus}. We also discussed \cite{prl} a singular field
dependence for the static magnetic moment in the underdoped
cuprates, and this is consistent with \cite{lake2,boris}.

The starting hypothesis of our theory is that the collective spin
excitations of the doped cuprates can be described by using the
proximity of a magnetic quantum critical point. This was proposed
in Ref. \cite{sy}; almost simultaneously, NMR experiments in
La$_{2-\delta}$Sr$_{\delta}$CuO$_4$ \cite{imai} showed crossovers
which could be neatly interpreted in terms of a magnetic quantum
critical point near a doping concentration $\delta \approx 0.12$
with dynamic exponent $z=1$. The ground state is a good
superconductor at this value of $\delta$, and so the magnetic
transition takes place between two superconducting states.
Evidence supporting this interpretation also appeared in neutron
scattering measurements \cite{aeppli}. An explicit theory for a
quantum transition between a $d$-wave superconductor with
co-existing long-range spin density wave order (a SC+SDW state)
and an ordinary $d$-wave superconductor (a SC state) was first
discussed by Balents {\em et al.} \cite{bfn}; they focused on the
case where the SDW ordering wavevector was exactly equal to the
spacing between the two nodal points where the $d$-wave
superconductor has gapless quasi-particle excitations, and studied
their role in the critical theory. However, their analysis also
makes it clear that the nodal quasiparticles can be safely
neglected for the generic case in which the wavevector matching
condition is not satisfied \cite{vs}, and we will mainly discuss
this simpler case here. The SC+SDW to SC transition in this case
is formally identical to that in an insulator, and the SC state
has a sharp $S=1$ `resonance peak' associated with stable $S=1$
collective excitonic excitation \cite{csy}. Such a theory for the
SC to SC+SDW transition was used by us \cite{sbv} to predict the
effects of Zn impurities on the resonance peak in the SC phase.

\section{Order parameter and field theory}
\begin{figure}
\centerline{\includegraphics[width=4in]{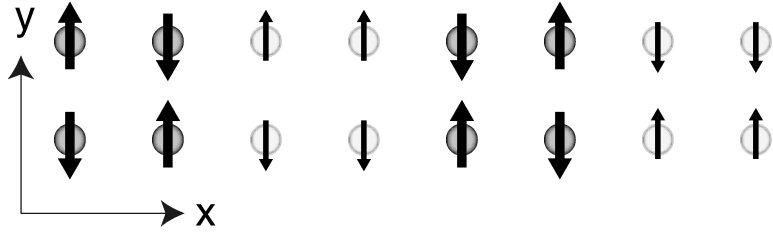}} \caption{A
bond-centered, collinear SDW at the wavevector $(3 \pi /4, \pi)$.
The size of the arrow represents the mean spin moment, while the
shading of the circle is the electron density. The sliding degree
of freedom corresponds to a shift in the position of the light and
dark circles.} \label{sdw}
\end{figure}

Neutron scattering experiments show that the lowest energy
collective spin excitations at and above $\delta \approx 0.12$
reside near the wavevectors ${\bf K}_x = (3 \pi/4, \pi)$ and ${\bf
K}_y = (\pi, 3 \pi/4)$ (our unit of length is the square lattice
spacing). So we write for the spin operator at the site ${\bf r}$:
\begin{equation}
S_{\alpha} ({\bf r}, \tau) = \mbox{Re} \left[e^{i {\bf K}_{x}
\cdot {\bf r}} \Phi_{x \alpha} ({\bf r}, \tau)+  e^{i {\bf K}_{y}
\cdot {\bf r}} \Phi_{y \alpha}({\bf r}, \tau)\right], \label{e1}
\end{equation}
where $\alpha=x,y,z$ extends over the directions in spin space,
and $\Phi_{x,y\alpha}$ are {\em complex} fields which will serve
as order parameters for the SC+SDW to SC transition. In the SC+SDW
phase, $\Phi_{x,y\alpha}$ are condensed and the condensate
describes the spatial modulation of the spin, by (\ref{e1}); in
the SC phase, the $\Phi_{x,y\alpha}$ are dynamically fluctuating,
and its quanta are {\em spin excitons}. The representation
(\ref{e1}) can describe a large variety of spin modulations {\em
e.g.} the state with $\langle \Phi_{x \alpha} \rangle \propto
(1,i,0)$, $\langle \Phi_{y \alpha} \rangle = 0$ is a spiral SDW in
the $x$ direction. Experimentally, however, it is clear that the
SDW is not spiral, but {\em collinear}; an example of a collinear
SDW has $\langle \Phi_{x \alpha} \rangle \propto e^{i \theta}
(1,0,0)$, $\langle \Phi_{y \alpha} \rangle = 0$---notice that the
average spin vectors on all sites are parallel or antiparallel.
The phase $\theta$ represents a {\em sliding} degree of freedom of
the SDW: for the special commensurate value of ${\bf K}_x$ under
consideration here, the coupling to the lattice will prefer that
$\theta$ take one of the values $n \pi/4$ (a site-centered SDW) or
$(n+1/2) \pi/4$ (a bond-centered SDW) where $n=0 \ldots 7$
integer. A sketch of a bond-centered SDW is shown in
Fig~\ref{sdw}.
Notice that the magnitude of the spin changes from site to site,
and this implies \cite{zachar} that there must be a corresponding
modulation in the electron density, $\delta \rho ({\bf r})$: the
latter has the same quantum numbers as $\sum_{\alpha}
S_{\alpha}^2$, and hence we deduce that the charge order can be
written as
\begin{equation}
\delta \rho ({\bf r}, \tau) \propto \mbox{Re} \sum_{\alpha}
\left[e^{i 2 {\bf K}_{x} \cdot {\bf r}} \Phi_{x \alpha}^2 ({\bf
r}, \tau)+ e^{i 2 {\bf K}_{y} \cdot {\bf r}} \Phi_{y \alpha}^2
({\bf r}, \tau)\right]+ \ldots. \label{e2}
\end{equation}
The period of the charge order is half that of the SDW, and the
amplitude of the charge order vanishes for the spiral SDW. We
emphasize that after accounting for the long-range Coulomb
interactions, the actual modulation in the electron charge density
per site may well be unobservably small. The ``charge order''
discussed here and in (\ref{e2}) should be interpreted in a much
more general sense, as representing a modulation at wavevectors $2
{\bf K}_{x,y}$ in all spin-singlet observables which are invariant
under time-reversal; the modulation could be larger in other
observables like the mean kinetic energy, exchange energy, or
pairing amplitude in the bonds between nearest-neighbor sites.

We are interested here in the quantum transition from the SC+SDW
state with $\langle \Phi_{x,y\alpha} \rangle \neq 0$ to the SC
state with $\langle \Phi_{x,y\alpha} \rangle = 0$ in a background
of quiescent superconductivity. The allowed terms in the effective
action are constrained by the underlying symmetries: these were
discussed in some generality in \cite{prb}. Here we will be
satisfied by considering a simplified effective action which is
written most easily in terms of the real and imaginary components
of $\Phi_{x,y\alpha}$:
\begin{equation}
\begin{array}{lll}
\Phi_{xx} = \varphi_1 + i \varphi_7 & \Phi_{xy} = \varphi_2 + i
\varphi_8 & \Phi_{xz} = \varphi_3 + i \varphi_9 \\
\Phi_{yx} = \varphi_4 + i \varphi_{10} & \Phi_{yy} = \varphi_5 + i
\varphi_{11} & \Phi_{yz} = \varphi_6 + i \varphi_{12}
\end{array} \label{ri}
\end{equation}
The effective action for the real fields $\varphi_{\mu}$, $\mu = 1
\ldots 12$ is taken to be
\begin{equation}
\mathcal{S}_{\varphi} = \int d^2 r  d \tau \Bigl\{ \frac{1}{2}
\Bigl[ (\partial_{\tau} \varphi_{\mu})^2 + ({\bf \nabla}_{\bf r}
\varphi_{\mu})^2 + s  \varphi_{\alpha}^2 \Bigr] + \frac{u}{2}
(\varphi_{\mu}^2)^2 \Bigr\} \label{e3}
\end{equation}
where a summation over the repeated $\mu$ index is implied, and we
have chosen units so that the velocity of spin waves is unity.
Notice that the action $\mathcal{S}_{\varphi}$ has a large O(12)
symmetry of rotations in $\mu$ space. This symmetry is present in
{\em all} allowed terms which are quadratic in $\varphi_{\mu}$,
but is broken by a number of quartic terms \cite{prb} which are
not displayed in (\ref{e3}). However, all permitted terms in the
effective action do respect the sliding symmetry under which
$\Phi_{x,y \alpha} \rightarrow e^{i n_{x,y} \pi/4} \Phi_{x,y
\alpha}$ for integer $n_{x,y}$. The coupling $s$ serves as the
tuning parameter which measures distance from the quantum critical
point: the SC+SDW phase will appear for $s<s_c$ and the
spin-singlet SC phase for $s>s_c$. The dynamic properties of the
transition at $s=s_c$ have been investigated in much detail in
earlier work \cite{csy}.

Now we consider the influence of the magnetic field, $H$, applied
perpendicular to the CuO$_2$ layers. This couples most strongly to
the `background' SC order, and so we are forced to consider the
response of the superconducting order parameter $\psi ({\bf r})$.
In suitable units (discussed in \cite{prb}), the free energy for
$\psi ({\bf r})$ can be written in the familiar Ginzburg-Landau
form
\begin{equation}
{\cal F} = \Upsilon \int d^2 r \left[- |\psi|^2 + \frac{1}{2}
|\psi|^4 + \left| \left( {\bf \nabla}_{\bf r} - i {\bf A} \right)
\psi \right|^2 \right] \label{f}.
\end{equation}
where $\Upsilon$ is a parameter measuring the relative
contributions of the magnetic and superconducting energies, and
$\nabla_{\bf r} \times {\bf A} = H \hat{\bf z}$. Notice that we
have assumed a $\tau$ independent $\psi$---this is permissible
because the SC order is non-critical and its quantum fluctuations
can be safely neglected.

Finally, we have to couple the $H$-response of $\psi ({\bf r})$ to
the quantum SDW fluctuations. There are two distinct couplings,
which have rather different physical consequences. The first, $v$,
is a simple coupling between the magnitudes of the SC and SDW
order parameters, chosen with a repulsive sign ($v>0$) to account
for the competition between these orders:
\begin{equation}
\mathcal{S}_v = \frac{v}{2} \int d^2 r d \tau \varphi_{\mu}^2
({\bf r}, \tau) |\psi ({\bf r})|^2. \label{v}
\end{equation}
Such a coupling was discussed by Zhang \cite{so5}, in work
focusing on the possibility of a {\em first order} transition
between an SC phase and an insulating phase with SDW order. The
coupling $v$ will play an important role in determining our phase
diagram to be described below, which contains a {\em second order}
transition between SC and SC+SDW phases. The second coupling,
unlike $v$, recognizes the fact that the vortex lattice induced by
$H$ breaks translational symmetry, and so the SDW fluctuations
should also not be invariant under the `sliding' symmetry (the
term in (\ref{v}) is invariant under the sliding symmetry). In
particular the vortex core radius in the cuprates is only of the
order of a few lattice spacings, and the energy of an SDW
fluctuation will certainly change depending upon which portion of
the charge order (see Fig~\ref{sdw}) is centered on a vortex core.
In other words, each vortex core, at a position ${\bf r}_v$, will
prefer a certain phase of the local charge order parameter
$\sum_{\alpha} \Phi_{x,y\alpha}^2 ({\bf r}_v, \tau)$. Expanding
the charge order parameter using (\ref{ri}), we deduce the second
term which couples the SC and SDW order parameters:
\begin{eqnarray}
&& \mathcal{S}_{\rm pin} = -\sum_{\{{\bf r}_v, \psi({\bf r}_v) =
0\}} \int d \tau \Biggl\{\sum_{\mu=1}^3 \mbox{Re} \biggl[
\zeta_{x} \left(\varphi_{\mu} ({\bf r}_v, \tau) + i
\varphi_{6+\mu} ({\bf
r}_v, \tau) \right)^2 \biggr]  \nonumber \\
&&~~~~~~~~~~~~~~~~~~~~~~~+\sum_{\mu=4}^6 \mbox{Re} \biggl[
\zeta_{y} \left(\varphi_{\mu} ({\bf r}_v, \tau) + i
\varphi_{6+\mu} ({\bf r}_v, \tau) \right)^2 \biggr] \Biggr\}.
\label{spin}
\end{eqnarray}
The complex coupling constants $\zeta_{x,y}$ measure the pinning
strength of the phase of the sliding charge order to some
preferred value near each vortex core.

We have now defined a well-posed field theoretical problem, which
was analyzed in some detail in \cite{prb}: describe the dynamic
quantum SDW fluctuations associated with the partition function
\begin{equation}
\mathcal{Z}\left[\psi ({\bf r}) \right] = \int {\cal D}
\varphi_{\mu} ({\bf r}, \tau) \exp \left( - \frac{{\cal F}}{T} -
{\cal S}_{\varphi} - {\cal S}_{v} - {\cal S}_{\rm pin} \right),
\label{zdef}
\end{equation}
where the optimum value of the static SC order $\psi ({\bf r})$ is
determined by the minimization of $-\ln \mathcal{Z}\left[\psi
({\bf r}) \right]$ via the solution of the saddle-point equation
\begin{equation}
\frac{\delta \ln \mathcal{Z}\left[\psi ({\bf r}) \right]}{\delta
\psi ({\bf r})} = 0. \label{scdef}
\end{equation}
Note the highly asymmetric treatment of the SC and SDW orders.

\section{Phase diagram}

\begin{figure}
\centerline{\includegraphics[width=3.5in]{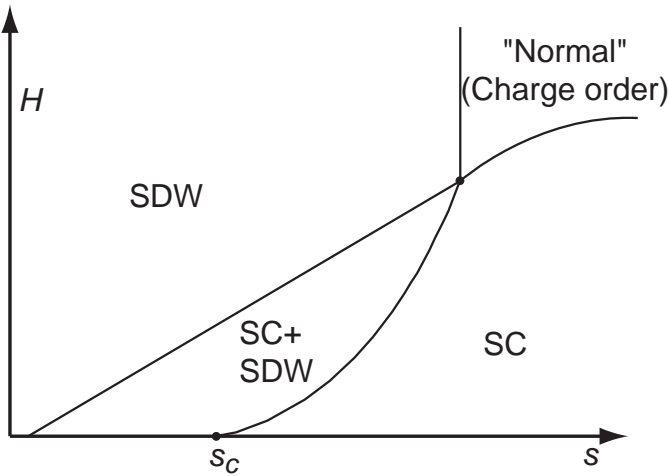}}
\caption{Zero temperature phase diagram of the model $\mathcal{Z}$
defined in (\protect\ref{zdef}) and (\protect\ref{scdef}); from
Refs.~\protect\cite{prl} and~\protect\cite{prb}. The phases have
the following expectation values: ({\em i\/}) SC: $\langle
\varphi_{\mu} \rangle = 0$, $\langle \psi \rangle \neq 0$, ({\em
ii\/}) SC+SDW: $\langle \varphi_{\mu} \rangle \neq 0$, $\langle
\psi \rangle \neq 0$, ({\em iii\/}) SDW: $\langle \varphi_{\mu}
\rangle \neq 0$, $\langle \psi \rangle = 0$, and ({\em iv\/})
``Normal'': $\langle \varphi_{\mu} \rangle = 0$, $\langle \psi
\rangle = 0$.} \label{figpd}
\end{figure}

Our primary results for the properties of (\ref{zdef}) and
(\ref{scdef}) are contained in the phase diagram as a function of
$s$ and $H$ in Fig~\ref{figpd}.
The formal solution of these equations also allows solutions in
which the SC order vanishes and $\psi ({\bf r} ) = 0$ everywhere:
this leads to the SDW phase and the ``Normal'' phase. However we
do not expect our theory to be accurate such a regime: quantum
fluctuations of the SC order parameter are surely important once
$\psi ({\bf r})$ becomes small, and these have been neglected in
our theory. Our results are more precise in the small $H$ region
of Fig~\ref{figpd}, in the vicinity of the boundary between the
SC+SDW and SC phases; indeed, the functional forms of the main
results quoted below are expected to be exact. Upon accounting for
the quantum fluctuations of the superconducting order, and the
Berry phases associated with the electrons in the nearby Mott
insulator \cite{vs}, we expect that the ``Normal'' phase will
display some sort of static charge order.

An important prediction of our theory is the shape of the
second-order phase boundary between the SC and SC+SDW phases at
small $H$. This transition is associated with the condensation of
the $\varphi_{\mu}$ exciton, when the spin gap to the creation of
an exciton vanishes in the SC phase. Detailed arguments were
presented in \cite{prl,prb} showing that this condensation occurs
in an exciton state which is extended throughout the entire
lattice; indeed, a variational approximation in which the exciton
wavefunction is assumed to be simply a constant gives essentially
exact results (as opposed to an approximation in which the exciton
is strongly localized in the vortex cores \cite{arovas}). The
presence of the vortex lattice in $\psi ({\bf r})$ influences the
energy of this extended exciton primarily via the $v$ coupling in
$\mathcal{S}_v$. When spatially averaged over ${\bf r}$, the
dominant change in the average value of $|\psi ({\bf r})|^2$, and
hence in the energy of the exciton, arises from slight suppression
of superconductivity in the {\em superflow\/} region surrounding
each vortex core. The much smaller vortex core region always has a
significantly weaker effect on the exciton energy. It is a simple
matter to compute the correction to the exciton energy from the
superflow. The average kinetic energy of the superflow is known
from standard Ginzburg-Landau theory to be $\sim (H/H_{c2}) \ln
(H_{c2}/H)$ (here $H_{c2}$ is the upper critical field at which
superconductivity disappears), and the coupling $v$ in (\ref{v})
therefore leads to a corresponding change in the effective value
of $s$ controlling the exciton energy:
\begin{equation}
s_{\rm eff} = s - \mathcal{C} \frac{H}{H_{c2}} \ln \left(
\frac{H_{c2}}{H} \right), \label{seff}
\end{equation}
where $\mathcal{C}$ is a positive constant. The critical field at
which the spin gap vanishes is therefore determined by $s_{\rm
eff} = s_c$, and this leads to our result \cite{prl,prb} for the
phase boundary between the SC and SC+SDW phases:
\begin{equation}
H \sim \frac{(s-s_c)}{\ln (1/(s-s_c))}. \label{hs}
\end{equation}
Note that the phase boundary approaches the $H=0$ limit with
vanishing slope: consequently a relatively small field applied to
the superconductor for $s>s_c$ will drive the system into the
SC+SDW phase. This is our explanation for the shift in the energy
of the dynamic spin fluctuations seen in \cite{lake1}.

The phase diagram of Fig~\ref{figpd} leads to a number of
predictions \cite{prl,kwon,prb} for observables in the SC and
SC+SDW phases, some of which have been tested in recent
experiments \cite{seamus,lake2,boris}. We discuss theory and
experiment in the two phases in turn in the following subsections.

\subsection{Static charge order in the SC phase}
\label{sec:sc}

The SC phase has $\langle \varphi_{\mu} \rangle = 0$, and so the
SDW fluctuations are dynamic and the spin exciton only exists
above a finite energy gap $\Delta$. The superconducting order
$\psi ({\bf r})$ is suppressed in the vortex cores, and so this
region should exhibit characteristics of the doped spin-gap ({\em
i.e.} paramagnetic) Mott insulator, as was argued in
\cite{ssvortex}. Paramagnetic Mott insulators, and their response
to doping with mobile charge carriers, were studied at some length
in \cite{vs}: it was argued that the most likely candidates had
bond-centered charge order which survived in a superconducting
state for a finite range of doping---this work will be reviewed
further in Section~\ref{sec:p4}. Reasoning in this manner,
Ref.~\cite{kwon} predicted that static charge order should appear
in and around the vortex cores, coexisting with dynamic spin
fluctuations in the SC state. Order of this type appears to have
been seen in the recent STM experiment on
Bi$_2$Sr$_2$CaCu$_2$O$_{8+\delta}$ \cite{seamus}. Our approach
should be contrasted from other recent investigations \cite{ting}
which have {\em both} static spin and charge order in the vortex
core.

\begin{figure}
\centerline{\includegraphics[width=3.5in]{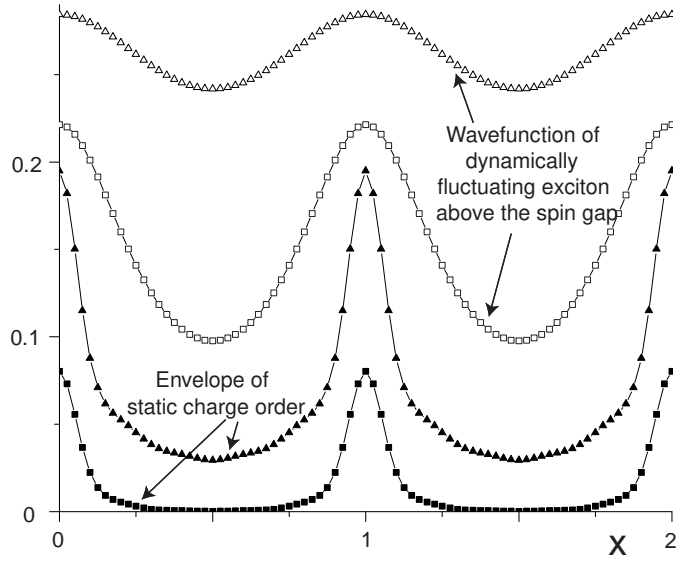}}
\caption{Spatial structure of the spin and charge correlations in
the SC phase at two different magnetic fields (squares and
triangles; the squares have the lower field). The vortices are
centered at $x=0,1,2$. The orientation of the spins fluctuates
with a frequency of order the inverse spin gap $\hbar/\Delta$. The
open symbols represent the lowest energy exciton wavefunction
$\Xi_{0 {\bf 0}} ({\bf r})$. The nature of the microscopic spin
correlations can be understood by recalling that this wavefunction
is the envelope of the order Fig~\protect\ref{sdw}. After
including the pinning term $\mathcal{S}_{\rm pin}$, static charge
order is induced, and its envelope $\Omega ({\bf r})$ (defined in
(\protect\ref{omega1},\protect\ref{omega2})) is shown above. The
numerical results were obtained by Ying Zhang and reported in
\protect\cite{prb}.} \label{figexciton}
\end{figure}

The spatial extent of the charge order was computed in the field
theory in (\ref{zdef}), (\ref{scdef}) in \cite{prb}. The gapped
spin exciton, $\varphi_{\mu}$, views the vortex lattice as a
periodic potential, and consequently its dispersion develops the
Bloch structure of a particle moving in a periodic potential. To
zeroth order in the pinning terms, $\zeta_{x,y}$, the two-point
$\phi_{\mu}$ Green's function is diagonal in the $\mu$ index, and
its diagonal component can be written as \cite{prl,prb}
\begin{equation}
G_{\varphi} ({\bf r}, {\bf r}', \omega_n) = \sum_{\mu} \int_{1BZ}
\frac{d^2 k}{4 \pi^2} \frac{\Xi_{\mu {\bf k}}^{\ast} ({\bf r})
\Xi_{\mu {\bf k}} ({\bf r}')}{\omega_n^2 + E_{\mu}^2 ({\bf k})},
\label{geigen}
\end{equation}
where ${\bf k}$ is a Bloch momentum which extends over the first
Brillouin zone of the vortex lattice, $\mu$ is a `band' index,
$\Xi_{\mu {\bf k}} ({\bf r})$ are the Bloch states ($\Xi_{\mu {\bf
k}} ({\bf r} + {\bf R}_v) = e^{i {\bf k} \cdot {\bf R}_v} \Xi_{\mu
{\bf k}} ({\bf r})$ where ${\bf R}_v$ is an vector connecting two
vortex centers), $E_{\mu} ({\bf k})$ are the energy dispersions of
the various bands, and $\omega_n$ is an imaginary Matsubara
frequency. The lowest energy exciton has energy $E_0 ({\bf 0})
\equiv \Delta$ and wavefunction $\Xi_{0 {\bf 0}} ({\bf r})$; this
wavefunction is sketched for typical parameter values in
Fig~\ref{figexciton}.

Now let us consider the influence of the pinning term in
(\ref{spin}). Combining (\ref{spin}) with (\ref{e2}) it is simple
to see that to first order in the $\zeta_{x,y}$, static charge
order appears in the SC phase, with
\begin{equation}
\langle \delta \rho ({\bf r}) \rangle \propto \mbox{Re} \left[
\zeta_{x} e^{i 2 {\bf K}_x \cdot {\bf r}} + \zeta_{y} e^{i 2 {\bf
K}_y \cdot {\bf r}} \right] \Omega ({\bf r}) \label{omega1}
\end{equation}
where
\begin{equation}
\Omega ({\bf r}) \equiv T \sum_{\omega_n} \sum_{{\bf r}_v}
G^2_{\varphi} ({\bf r}, {\bf r}_v, \omega_n). \label{omega2}
\end{equation}
A plot of the function $\Omega ({\bf r})$ is sketched in
Fig~\ref{figexciton}, along with the corresponding $\Xi_{0 {\bf
0}} ({\bf r})$. At the higher field (triangles), the spin exciton
wavefunction $\Xi_{0{\bf 0}} ({\bf r})$ is essentially constant
across the entire system, with only a weak modulation induced by
the vortex lattice; nevertheless, at the same field the charge
order $\Omega ({\bf r})$ has a strong modulation on the scale of
$c/(2\Delta)$, where $c$ is a spin-wave velocity \cite{pphmf}. At
the lower field (squares), there is larger modulation in the spin
exciton (but $\Xi_{0 {\bf 0}} ({\bf r})$ only decays to half its
maximum value), and again the decay length of the charge
correlations is about half that of the spin correlations. The
spatial form of the lower field $\Omega ({\bf r})$ in
Fig~\ref{figexciton} is quite similar to envelope of the
modulation observed in \cite{seamus}.

The STM experiments of \cite{seamus} actually measure the
modulation in the local electronic density of states (LDOS) in a
range of energies as a function position in vortex lattice. A
great deal of information is, in principle, contained in the
energy and spatial dependence of the LDOS modulations. We have
recently analyzed \cite{pphmf,rc} simple models for the coupling
of the electronic quasiparticles to the collective SDW and SC
degrees of freedom that have been discussed here: these lead to
predictions for the LDOS modulations, which can be usefully
compared with the STM data---the reader is referred to the papers
for details.

We also mention here the recent STM experiments of Howald {\em et
al.} \cite{aharon} which have observed charge order similar to
that of \cite{seamus} but in zero applied magnetic field; this
order has (presumably) been pinned by impurities.

\subsection{Static spin moment in the SC+SDW phase}

The SC+SDW phase has $\langle \varphi_{\mu} \rangle \neq 0$, and
hence via (\ref{e1}), (\ref{e2}), and (\ref{ri}), there is both
static spin and charge order. Neutron scattering measurements have
so far only succeeded in observing the static spin order, and so
we will restrict our discussion here to the spin moment.

In the presence of an applied magnetic field, the vortex lattice
will spatially modulate the value of $\langle \phi_{\mu} ({\bf r})
\rangle $, and this should, in principle, lead to satellite
elastic peaks \cite{arovas,prl,prb} surrounding the main elastic
Bragg peaks at $\pm {\bf K}_x$, $\pm {\bf K}_y$ observed in
neutron scattering. However, our discussion above on the dominance
of the superflow effect shows that this modulation occurs
predominantly on the scale of the vortex lattice spacing
\cite{prl,prb}, and not on the scale of the vortex core. The
resulting satellite peaks are consequently found to be extremely
weak, as will be clear from our results below.

\begin{figure}
\centerline{\includegraphics[width=3.5in]{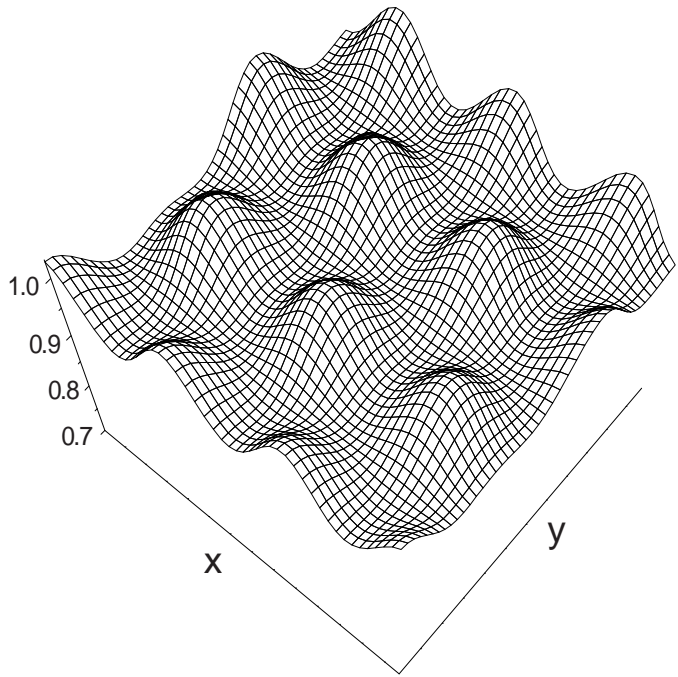}}
\caption{Spatial form of $\langle \varphi_{\mu} ({\bf r})$ in the
SC+SDW phase. Notice the vertical scale, which shows that the
overall modulation in the size of the order parameter is quite
small. The numerical results were obtained by Ying Zhang and
reported in \protect\cite{prb}.} \label{figsurface}
\end{figure}

\begin{figure}
\centerline{\includegraphics[width=5.5in]{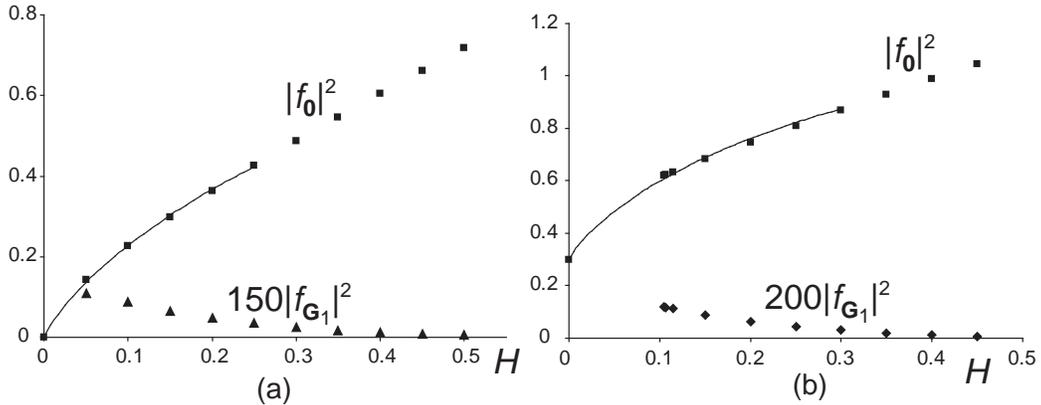}}
\caption{Magnitudes of the elastic scattering peaks in
(\protect\ref{sko}) obtained by the spatial Fourier transform of
results like those in Fig~\protect\ref{figsurface}. The field $H$
is measured in units of $H_{c2}^0$, the value of the critical
field at which all four phases meet in Fig~\protect\ref{figpd}.
The plot (a) is at $s=s_c$, while (b) is for $s<s_c$. All other
couplings are identical to those in Figs~\protect\ref{figexciton}
and~\protect\ref{figsurface}. The numerical results were obtained
by Ying Zhang and reported in \protect\cite{prb}.} \label{elastic}
\end{figure}

We show a sketch of the spatial form of $\langle \varphi_{\mu}
({\bf r})$ at a point in the SC+SDW phase in Fig~\ref{figsurface}.
All the coupling constants in the theory are the same as those
used for the results in Fig~\ref{figexciton} for the SC phase.
Only the parameters $s$ and $H$ are tuned to move the system
between the SC and SC+SDW phases.
The spatial Fourier transform of Fig~\ref{figsurface} determines
the strength of the elastic Bragg peaks that will be observed in
neutron scattering. In particular, the dynamic structure factor of
$\varphi_{\mu}$ has the form
\begin{equation}
S_{\varphi} ({\bf k}, \omega) = (2 \pi) \delta (\omega) \sum_{\bf
G} |f_{\bf G}|^2 (2 \pi)^2 \delta({\bf k} - {\bf G}) \label{sko}
\end{equation}
where ${\bf G}$ are the reciprocal lattice vectors of the vortex
lattice. The reader should keep in mind, via (\ref{e1}) and
(\ref{ri}), that the experimental structure factor is obtained
from (\ref{sko}) by measuring wavevectors from the SDW ordering
wavevectors $\pm {\bf K}_x$, $\pm {\bf K}_y$.  We show typical
results for the field dependence of the strength of the central
peak, $|f_{{\bf 0}}|^2$, and also for the first satellite peak,
$|f_{{\bf G}_1}|^2$ (where ${\bf G}_1$ is the smallest non-zero
reciprocal lattice vector) in Fig~\ref{elastic}.
The strong observable effect is in the $H$ dependence of the
central peak $|f_{{\bf 0}}|^2$. The same superflow effects which
were responsible for (\ref{hs}), also dominate in determining the
average magnitude of the SDW order in the SC+SDW phase: we can
estimate the enhancement of magnetic order by the superflow by
assuming that $f_{{\bf 0}}$ is determined by $s_{eff}$, and then
(\ref{seff}) leads to
\begin{equation}
|f_{\bf 0}|^2 (H) - |f_{\bf 0}|^2 (0) \propto H \ln (1/H).
\label{hln}
\end{equation}
the lines in Fig~\ref{elastic} are fits of the full numerical
solution to (\ref{hln}). Arovas {\em et al.} \cite{arovas} had
discussed nucleation of static magnetic order in the vortex core
in what was an SC phase. In contrast, we claim that there are {\em
no static spins in the vortices in the SC phase}, and only pinned
static charge order around each vortex as discussed in
Section~\ref{sec:sc}. The static moments appear only when there is
bulk magnetic order as in the SC+SDW phase (as pointed out in
\cite{prl,prb}), and here it is the contribution of the superflow
which always dominates leading to (\ref{hln}). The linear small
$H$ dependence of the ordered moment proposed by Arovas {\em et
al.} is not valid in either the SC or the SC+SDW phases.

The results in Fig~\ref{elastic} also compare well with the
experimental observations in the overall scale of both the field
and the magnetic moment: those in (a) are quite similar to the
results of \cite{lake2}, while (b) matches well with \cite{boris}.

Finally, note that Fig~\ref{elastic} shows that the satellite
peaks are unobservably small. This is related to relatively slow
modulation induced by the superflow in the moment in
Fig~\ref{figsurface}. Ref.~\cite{prb} predicted that the influence
of the vortex lattice may be more easily observable in the {\em
dynamic} exciton band structure in the SC phase.

\section{Why does the charge order have period 4 ?}
\label{sec:p4}

The SDW (${\bf K}_{x,y}$) and charge ($2 {\bf K}_{x,y}$) ordering
wavevectors have so far been arbitrary parameters in our
phenomenological theory, and the structure of this theory is
largely independent of the values of ${\bf K}_{x,y}$ (some high
order terms in the action are permitted only for certain
commensurate values of ${\bf K}_{x,y}$). The determination of
${\bf K}_{x,y}$ requires, instead, a lattice scale theory of the
doped antiferromagnet.

\begin{figure}[h]
\centerline{\includegraphics[width=3.5in]{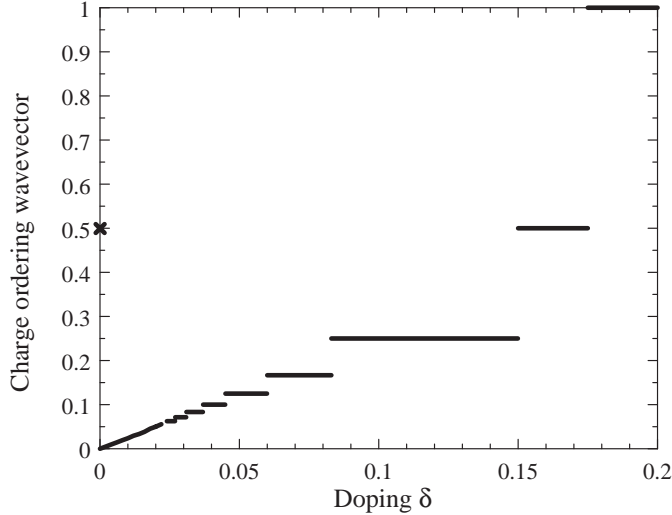}}
\caption{Evolution of the charge ordering wavevector upon doping a
paramagnetic Mott insulator, from Ref.~\cite{vs}. The wavevector
is measured in units of $2\pi/a$. At $\delta=0$, the Mott
insulator has period 2, which is associated with the appearance of
bond order (see Fig~\protect\ref{figphase}). Beyond a range of
very small $\delta$ values, the ground state is also a
superconductor. Full square lattice symmetry is restored above
$\delta \approx 0.175$, when the ground state becomes an ordinary
$d$-wave superconductor.} \label{fig:dop}
\end{figure}

The mechanism of the charge-ordering instability in doped
antiferromagnets has been discussed using a number of different
theoretical perspectives by several other workers
\cite{white,jan,steve,troyer,nayak}; in general, many possible
charge-ordering periods emerge in their works, and there appear to
be no fundamental principles restricting the possible periods, or
whether the charge-ordering is site or bond centered. Here, we
will briefly recall our theoretical work on charge ordering
\cite{vs}: its point of departure is the theory of the magnetic
quantum critical point in the Mott insulator. Our perspective led
to bond-centered charge ordered states, without long-range
magnetic order, and with co-existing $d$-wave-like
superconductivity. The predicted evolution of the wavevector of
this ordering, as a function of hole concentration, is shown in
Fig~\ref{fig:dop}.
Note that the period, $p$, is always pinned to be an even number
{\em i.e.} $2{\bf K}_x = (2 \pi/a)(1/p, 0)$, where $a$ is the
square lattice spacing. There is a large range of $\delta$ values
in Fig~\ref{fig:dop} where the period is pinned at $p=4$: this
corresponds to the values of $2{\bf K}_{x,y}$ observed in the STM
experiments \cite{seamus,aharon}. Both experiments also see the
modulation appearing simultaneously at $2{\bf K}_x$ and at $2 {\bf
K}_y$, leading to a checkerboard appearance in real space. Such
checkerboard patterns have been considered previously for the
$p=2$ case \cite{rs,dk,book,aa}, and were found to have an energy
very close to that of the state with charge order only along a
single direction; we can expect that a similar result applies for
$p=4$.

Further support for Fig~\ref{fig:dop} has emerged from recent
neutron scattering measurements of Mook {\em et al.} \cite{mook}
in the under-doped superconductor YBa$_2$Cu$_3$O$_{6.35}$: they
observed static charge order, and dynamic spin correlations, with
the the charge order period pinned rather precisely at $p=8$. This
is in accord with the plateau at $p=8$ over a range of small
$\delta$ values in Fig~\ref{fig:dop}, and should be contrasted
with the continuous evolution of ordering wavevector with $\delta$
which is usually assumed in the ``Yamada plot'' \cite{tran,yamada}
for periods larger than $p=4$.

\begin{figure}
\centerline{\includegraphics[width=4.5in]{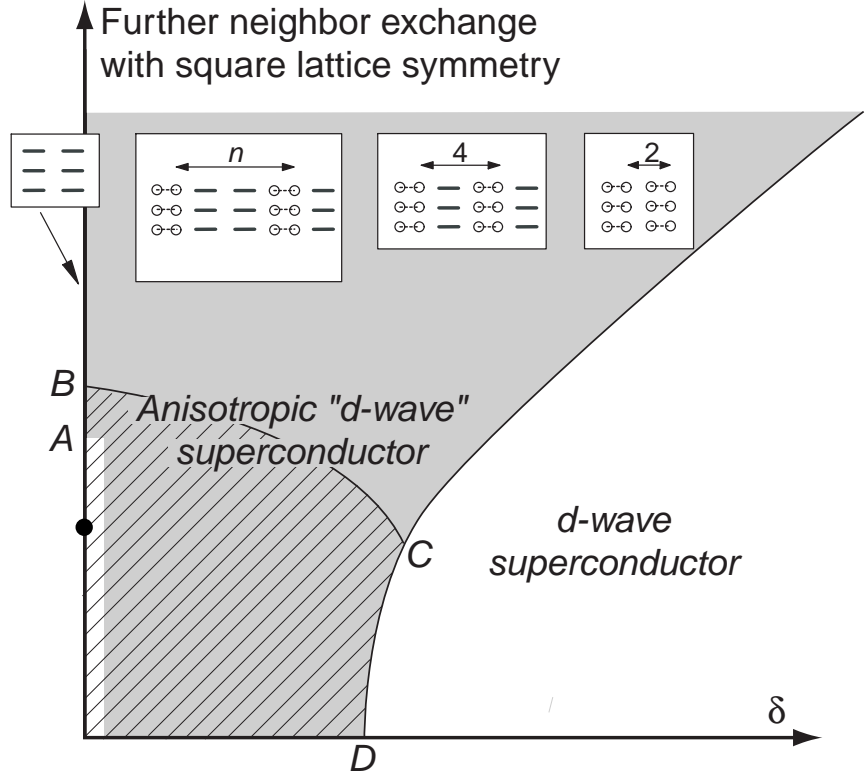}}
\caption{Schematic $T=0$ phase diagram (from
Refs.~\protect\cite{vs,sns}) for the high temperature
superconductors as a function of a ratio of the near neighbor
exchange interactions and the hole concentration, $\delta$; {\em
e.g.\/} the vertical axis could be $J_2/J_1$, the ratio of the
first to second neighbor exchange. The shaded region has charge
order. The hatched region has broken spin-rotation symmetry with
$\langle \vec{S} \rangle \neq 0$, and at least one of the order
parameters $\Phi_{x,y\alpha}$ is non-zero; $\langle \vec{S}
\rangle =0$ elsewhere. The unfrustrated, insulating
antiferromagnet with long-range N\'{e}el order is indicated by the
filled circle. At $\delta=0$, there is an onset of charge order
above the point A, while spin-rotation invariance is restored
above B. The nature of the charge orders as determined by the
computations of Ref.~\protect\cite{vs} are indicated at the top of
the figure; numerous transitions, within the gray shaded region,
in the nature of the charge ordering are not shown. The ground
state at very low non-zero doping is an insulating Wigner crystal
and there is subsequently a insulator-to-superconductor
transition; superconductivity is present over the bulk of the
$\delta >0$ region. The central idea behind our approach is that
many essential aspects the spin excitation spectrum of the
insulating, paramagnetic region ($\delta=0$, $\langle
\vec{S}\rangle =0 $) lead to a simple and natural description of
the analogous properties of the $d$-wave superconductor.}
\label{figphase}
\end{figure}

We conclude this paper by briefly recalling the physical
ingredients behind the results in Fig~\ref{fig:dop}. A review of
these arguments has already been presented in \cite{sns}, and we
present here a synopsis in Fig~\ref{figphase}.
In moving from an undoped Mott insulator, like La$_2$CuO$_4$, to a
high temperature superconductor at optimal doping, at least two
quantum phase transitions must take place: one involving the loss
of magnetic order, and the other the onset of superconductivity.
Theoretically, it is very useful to disentangle the two
transitions by imagining that we have a second tuning parameter at
our disposal, in addition to the doping $\delta$. We use this
second parameter to first destroy the magnetic order in
La$_2$CuO$_4$ while remaining at $\delta=0$---a specific
possibility for such a parameter is a frustrating second neighbor
exchange interaction (see Fig~\ref{figphase}). Detailed arguments
have been given (for a recent review see \cite{annals}) that the
paramagnetic Mott insulator so obtained has bond-centered charge
order, as indicated in Figs~\ref{fig:dop} and \ref{figphase}. The
second theoretical step of doping the paramagnetic Mott insulator
can be reliably addressed in a large $N$ theory \cite{vs}, and one
eventually obtains a $d$-wave superconductor which fully respects
the symmetries of spin rotations and lattice translations and
rotations. In this approach, the charge order of the
superconductor without magnetic long range order evolves from that
in the paramagnetic Mott insulator. Further, the Cooper pairing in
the superconductor is also connected to the singlet electron
pairing present in the bond-ordered paramagnetic Mott insulator.
These connections led us to the results in Fig~\ref{fig:dop}, and
to the proposal of charge order nucleation by vortices in
\cite{kwon}.

\begin{acknowledgments} The results reviewed here were obtained
with Eugene Demler, Kwon Park, Anatoli Polkovnikov, Matthias
Vojta, and Ying Zhang; I thank them for fruitful collaborations.
This research was supported by US NSF Grant DMR 0098226.
\end{acknowledgments}

\end{document}